\documentclass{article} 
\usepackage{graphicx}
\usepackage{amsmath}
\usepackage{amsfonts}
\usepackage{amssymb}
\usepackage{amscd}
\usepackage{graphics}
\usepackage{color}

\def\sumint{\sum\mspace{-25mu}\int}

\begin{document}

\title{Cluster properties in Poincar\'e invariant quantum mechanics 
\footnote{This work was supported in part by the U.S. Department 
of Energy, under 
contract DE-FG02-86ER40286; 
Contribution to the 20th International IUPAP Conference on Few-Body 
Problems in Physics, 
20 - 25 August, 2012, Fukuoka, Japan
}}
\author{
W. N. Polyzou \\ 
Department of Physics and Astronomy,\\
The University of Iowa, Iowa City, IA 52242\\
B. D. Keister \\
Physics Division, National Science Foundation,\\ 
Arlington, VA 22230
}


\maketitle

\begin{abstract}
Using a simple model we provide a quantitative study of the size of
the corrections needed to restore cluster properties to the
construction of Poincar\'e invariant dynamical models with kinematic
spins, first provided by B. Bakamjian and L. H. Thomas.  Our model
calculations suggest that these corrections are too small to have a
quantitative impact on nuclear physics observables calculated using
models with meson and nucleon degrees of freedom.

\end{abstract}

\section{Introduction}
\label{intro}

We provide a quantitative evaluation of the size of the operators that
restore cluster properties in Bakamjian-Thomas formulations of
relativistic few-body quantum mechanics.

Relativistic few-body models are an extension of the corresponding
non-relativistic models that are exactly Poincar\'e invariant.  By
exact Poincar\'e invariance we mean that quantum probabilities, which
are the dimensionless observables of the theory, have the same values
in all inertial coordinate systems.  Wigner\cite{wigner} showed that a
necessary and sufficient condition for the invariance of quantum
probabilities is the existence of a dynamical representation of the
Poincar\'e group on the model Hilbert space.

Dirac\cite{dirac} showed that at least three of the Poincar\'e
generators must have interactions in order to satisfy the commutation
relations in an interacting theory.  This is because time translation
can be expressed in terms of Lorentz boosts and spatial translations.

Beyond Poincar\'e invariance, the requirement that Poincar\'e
invariance also holds for isolated subsystems requires that the
unitary representation of the Poincar\'e group be well approximated by
a tensor product of two representations when evaluated between states
representing asymptotically separated subsystems.

Bakamjian and Thomas\cite{bakamjian} provided the first non-field
theoretic realization of the Poincar\'e Lie algebra with interactions.
Their construction satisfied cluster properties at the two-body level,
but not for systems of more than two particles.  Coester\cite{coester}
applied the Bakamjian-Thomas construction to three-body systems and
showed that the resulting three-body $S$-matrix satisfied cluster
properties.  Unfortunately his result did not extend to the unitary
representation of the Poincar\'e group and did not apply to systems of
four or more particles. Sokolov\cite{sokolov}\cite{coester2} provided
a complete solution to the problem in terms of certain unitary
operators.  In this framework the size of the corrections that restore
cluster properties are related to how close these unitary operators
are to the identity.  Sokolov's operators have never been computed in
any applications.

The Bakamjian-Thomas construction begins with a tensor product of two
irreducible representations of the Poincar\'e group and decomposes it
into a direct integral of irreducible representations using
Clebsch-Gordan coefficients for the Poincar\'e group
\[
\vert (M, j )\mathbf{P} ,\mu ; l ,s \rangle = 
\sumint d\mathbf{p}_1 d\mathbf{p}_2 
\vert (m_1,j_1) \mathbf{p}_1, \mu_1 \rangle \otimes \vert (m_2, j_2 )
\mathbf{p}_2, \mu_2 \rangle \times 
\]
\[
\underbrace{\langle   
(m_1,j_1) \mathbf{p}_1, \mu_1 (m_2, j_2 )\mathbf{p}_2, \mu_2 \vert 
(M, j )\mathbf{P} ,\mu ; l ,s \rangle}_{\mbox{Poincar\'e group Clebsch-Gordan coeff.}}.
\]
Interactions are added to the invariant mass operator of this free-particle
irreducible representation that commute with the free spin operator and commute 
with and are independent of the quantum numbers that label vectors in 
each irreducible subspace
\[
M_I = M_0 + V. 
\]
In the free-particle irreducible basis these matrix elements have the 
form
\[
\langle (M, j )\mathbf{P} ,\mu ; l ,s \vert V 
\vert (M', j' )\mathbf{P}' ,\mu' ; l' ,s' \rangle =
\]
\[
\delta(\mathbf{P}-\mathbf{P}')\delta_{jj'} \delta _{\mu \mu'} 
\langle M , l ,s \Vert v^j \Vert M',l',s' \rangle .
\]
Simultaneous eigenstates of $M_I, \mathbf{j}^2, j_z, \mathbf{P}$ are
constructed by solving the Schr\"odinger equation for the mass
eigenstates
\[
(\lambda -M) \phi_{\lambda ,j}( M,l,s)  =
\sumint' dM'
\langle M , l ,s \Vert v^j \Vert M',l',s' \rangle
\phi_{\lambda ,j}( M,l,s) .
\]
The eigenstates defined by the wave functions 
\[
\langle (M, j )\mathbf{P} ,\mu ; l ,s 
\vert (\lambda, j' )\mathbf{P}' ,\mu' \rangle =
\delta (\mathbf{P}-\mathbf{P}') \delta_{jj'}\delta_{\mu \mu'} 
\phi_{\lambda ,j}( M,l,s) 
\]
are complete and transform irreducibly under the dynamical representation
\[
U(\Lambda ,a ) \vert (\lambda, j )\mathbf{P} ,\mu \rangle
\]
\[
\sum_{\nu} \vert (\lambda, j )\pmb{\Lambda}{P} ,\nu \rangle
e^{i \Lambda P \cdot a} \sqrt{{\omega_\lambda (\Lambda P) 
\over \omega_\lambda (P) }}
D^j _{\nu \mu}(R_w(\Lambda ,P)) 
\]
of the Poincar\'e group.  The original Bakamjian-Thomas construction
was for a system of two particles; but a generalization of the
construction outlined above works for any number of particles.  The
key requirement is that the spin in the interacting model is
identified with the spin in the non-interacting model.

In the two-body Bakamjian-Thomas representation it is clear that when
the two-body interaction is turned off the resulting unitary
representation of the Poincar\'e group becomes the tensor product of
two non-interacting irreducible representations, as expected.  When
the Bakamjian-Thomas construction is applied to systems of three
particles, turning off the two-body interactions involving one
particle no longer results in a tensor product of a one and two body
representation of the Poincar\'e group.

The Sokolov construction starts with the two-body interactions that
appear in the two-body problem and uses them to construct three-body
interactions that lead to a dynamical representation of the Poincar\'e
group that clusters to the tensor product of the two-body
Bakamjian-Thomas representation and a one-body representation.  This
construction can be repeated recursively for any number of particles.

To understand the Sokolov construction consider a three-body system
where one pair of particles interact.  There are two natural
constructions of a dynamical representation of the Poincar\'e group.
The first is to take the tensor product of a two-body Bakamjian-Thomas
representation with a single-particle irreducible representation.  The
second is to perform a full three-body Bakamjian Thomas construction where
the interaction commutes with the non-interacting three-body spin.
Using appropriate choices of two-body interactions, these
constructions can be done in a manner that ensures that both
representations lead to the same scattering matrix elements and two
body-bound state masses.  The relevant additions to the two-body
invariant mass have the forms
\[
\langle \mathbf{P}',j_3',\mu_3', \mathbf{p}_3', m_{12}' ,j', l',s',\mu' 
\vert V^{TP}_{12} \vert 
\mathbf{P},j_3',\mu_3', \mathbf{p}_3, m_{12} ,j, l,s,\mu
\rangle =
\]
\[
\delta (\mathbf{P}' -\mathbf{P}) \delta_{j_3'j_3} 
\delta_{{\mu_3' \mu_3}}
\delta ({\mathbf{p}_3'} -{\mathbf{p}_3})
\delta_{j'j} \delta_{{\mu' \mu}}
{\langle m_{12}', l',s'\Vert v^j_{12} \Vert m_{12} ,l,s \rangle}
\]
and 
\[
\langle \mathbf{P}',{j}_3,\bar{\mu}_3, \mathbf{q}_3', m_{12}' ,{j}', l',s',
\bar{\mu}' 
\vert V^{BT}_{12} \vert 
\mathbf{P},{j}_3,\bar{\mu}_3, \mathbf{q}_3, m_{12} ,{j}, l,s,\bar{\mu}
\rangle =
\]
\[
\delta (\mathbf{P}' -\mathbf{P})\delta_{j_3'j_3} 
\delta_{{\bar{\mu}_3' \bar{\mu}_3}}
\delta ({\mathbf{q}_3'} -{\mathbf{q}_3})
\delta_{{j}'{j}} \delta_{{\bar{\mu}' \bar{\mu}}}
{\langle m_{12}', l',s'\Vert v^{j}_{12} \Vert m_{12} ,l,s \rangle}
\]
where 
\[
\mathbf{q}_3 := \pmb{\Lambda}^{-1} (P/M({k})) p_3  \qquad 
\bar{\mathbf{j}}= R_w(P,p_{12})\mathbf{j} \qquad 
\bar{\mathbf{j}}_3= R_w(P,p_{3})\mathbf{j}_3 .
\] 
Violations of cluster properties arise because the boost that appears
in the definition of $\mathbf{q}_3$ and $\bar{\mathbf{j}}$ depends on
$m_{12}$ which does not commute with the potential.  The spectator
delta functions and spin kronecker delta functions in these two
expressions are only equivalent when $m_{12}=m_{12}'$.  The $S$-matrix
in both for these representations are equal when the reduced kernels
$\langle m_{12}', l',s'\Vert v^{j}_{12} \Vert m_{12} ,l,s \rangle$ are
identified.  This is because the $S$-matrix has the form of the
reduced kernel
\[
\langle m_{12}', l',s'\Vert S^j_{12} \Vert m_{12} ,l,s \rangle
\]
multiplied by delta functions that become equivalent on shell (when
$m_{12}=m_{12}'$).  A consequence of this equivalence is that both
unitary representations of the Poincar\'e group are related by an
$S$-matrix preserving unitary transformation \cite{ekstein}
\[
A_{12,3} U^{TP}_{12,3}(\Lambda ,a) 
A_{12,3}^{\dagger} = U_{12,3}^{BT}(\Lambda ,a) .
\]
In the Bakamjian-Thomas representation it is possible to combine the
three $2+1$ mass operators to get an interacting three-body mass
operator that commutes with the non-interacting three-body spin
\[
M^{BT}:= M^{BT}_{12,3} + M^{BT}_{23,1} + M^{BT}_{31,2} - 2 M^{BT}_{0}
\qquad 
\mathbf{j}^{BT} := \mathbf{j}_0. 
\]
Applying the Bakamjian-Thomas construction to this mass operator gives
a dynamical unitary representation, $U^{BT} (\Lambda ,a)$, of the
Poincar\'e group.

Sokolov defined an $S$-matrix equivalent representation using a 
unitary transformation constructed from a symmetrized product of the 
three $2+1$ unitary operators relating the $2+1$ Bakamjian-Thomas 
representation to the $2+1$ tensor product representations: 
\begin{equation}
U(\Lambda ,a) := A^{\dagger} U^{BT} (\Lambda ,a)  A
\label{e.1}
\end{equation}
where
\[
A = e^{\mbox{ln}(A_{12,3})+\mbox{ln}(A_{23,1})+\mbox{ln}(A_{31,2})} .
\]
The resulting three-body invariant mass operator can be expressed
in terms of these unitary transformations and the mass operators
for the $2+1$ tensor product representations:
\[
M:= 
\]
\[
A^{\dagger} \left ( A_{12,3} M_{12,3} A^{\dagger}_{12,3}+ 
A_{23,1} M_{23,1}A^{\dagger}_{23,1} +
A_{31,2} M_{31,2}A^{\dagger}_{31,2} - 2 M^{BT}_{0} 
\right ) A .
\]
Because the $A_{ij,k} \to I$ when the interaction between particles
$i$ and $j$ is turned off, in each asymptotic region this mass
operator becomes the mass operator for the associated tensor product
representation, which implies that this transformed representation of
the Poincar\'e group satisfies cluster properties. The combined effect
of the unitary operators is to generate three-body interactions that
restore the Poincar\'e commutation relations to the cluster expansions
of the Poincar\'e generators.

In the limit that $A$ in (\ref{e.1}) becomes the identity the Sokolov
representation becomes the Bakamjian-Thomas representation.  Thus the
size of the difference between these unitary transformations and the
identity provides a measure of the size of the violations of cluster
properties in the Bakamjian-Thomas representation.

To test the size of the corrections that restore cluster properties we
consider a simple four-body model.  It consists of a three-particle
system where two of the particles interact to form a bound state and
an external probe that interacts weakly with the third particle.  We
assume that there are no interactions between the particles in the
bound pair and the third particle or the probe.  All particles are
treated as spinless particles and the probe is a assumed to interact
via a scalar ``current''.  We use a relativistic Malfliet-Tjon type of
potential to construct a model with nuclear-physics scales.

We formulate models treating the three-body system as a $2+1$
tensor-product representation or a $2+1$ Bakamjian-Thomas
representation.  The current matrix elements in the two cases are
related by the unitary transformation $A_{12,3}$
\[
\langle 12\otimes 3 \vert j(0) \vert 12\otimes 3' \rangle =
\langle 
(12,3)^{BT} \vert A_{12,3}j(0)A^{\dagger}_{12,3} 
\vert (12,3)^{\prime BT} \rangle  .
\]
It follows that the difference between 
$\langle 12\otimes 3 \vert j(0) \vert 12\otimes 3' \rangle$ 
and $\langle 
(12,3)^{BT} \vert j(0)  
\vert (12,3)^{\prime BT} \rangle$ provides one measure of difference 
between $A_{12,3}$ and the identity, which is a measure of the
size of the operator that restores cluster properties to the 
Bakamjian-Thomas representation.

In the figures we plot 
\[
F(\mathbf{p}_3'-\mathbf{p}_3, \mathbf{p}_{12}) :=
\]
\begin{equation}
{
\int d\mathbf{p}_{12}'{}_{TP}\langle  \mathbf{p}_3, 
\mathbf{p}_{12}, \phi \vert j(0) \vert  
\mathbf{p}_3', \mathbf{p}_{12}',\phi \rangle_{TP} -
\int d\mathbf{p}_{12}'{}_{BT}\langle \mathbf{p}_3, 
\mathbf{p}_{12}, \phi \vert j(0) \vert  
\mathbf{p}_3', \mathbf{p}_{12}',\phi \rangle_{BT} 
\over 
\int d\mathbf{p}_{12}'{}_{TP}\langle  \mathbf{p}_3, 
\mathbf{p}_{12}, \phi \vert j(0) \vert  
\mathbf{p}_3', \mathbf{p}_{12}',\phi \rangle_{TP}
}
\label{eq.2}
\end{equation}
In (\ref{eq.2}) the integral over $\mathbf{p}_{12}'$ removes the
dependence on the momentum of the bound pair, $\mathbf{p}_{12}$, in a
model that satisfies cluster properties.  $F(\mathbf{q}, \mathbf{p}_{12})$ 
must vanish for
models satisfying cluster properties, which is illustrated by the flat
plane in each of the figures.  Any residual dependence on
$\mathbf{p}_{12}$ in this expression indicates a violation of cluster
properties, which provides a measure of how much the operator
$A_{12,3}$ differs from the identity.  Figures 1. and 2. show
(\ref{eq.2}) for $p_{12}$ perpendicular and parallel to $q=p_3'-p_3$
in Dirac's front-form dynamics.  These two plots exhibit a small
dependence on $\mathbf{p}_{12}$, but the value differs from zero.
Figures 3. and 4. show (\ref{eq.2}) for $p_{12}$ perpendicular and
parallel to $q=p_3'-p_3$ in Dirac's instant-form dynamics.  Figures
5. and 6. show (\ref{eq.2}) for $p_{12}$ perpendicular and parallel to
$q=p_3'-p_3$ in Dirac's point-form dynamics.  Both the instant and
point-form calculations have more dependence on $\mathbf{p}_{12}$ than
the front-form calculation, but the magnitude of the violations of
cluster properties are of comparable size in all three cases.  While
all six plots exhibit clear violations of cluster properties, the size
of the violations are a few parts in a thousand which is well within
the size of both theoretical and experimental uncertainties in
relativistic nuclear physics observables.  The violations of cluster
properties increase with stronger binding or for wave functions with
higher mean momentum, however for scales associated with realistic
nuclear-nucleon interactions the violations remain small.  This
suggests that at current levels of experimental precision there is no
real need to compute corrections that restore cluster properties.

\begin{figure}
\begin{minipage}[t]{6.2cm}
\begin{center}
\includegraphics[width=5.9cm,clip]{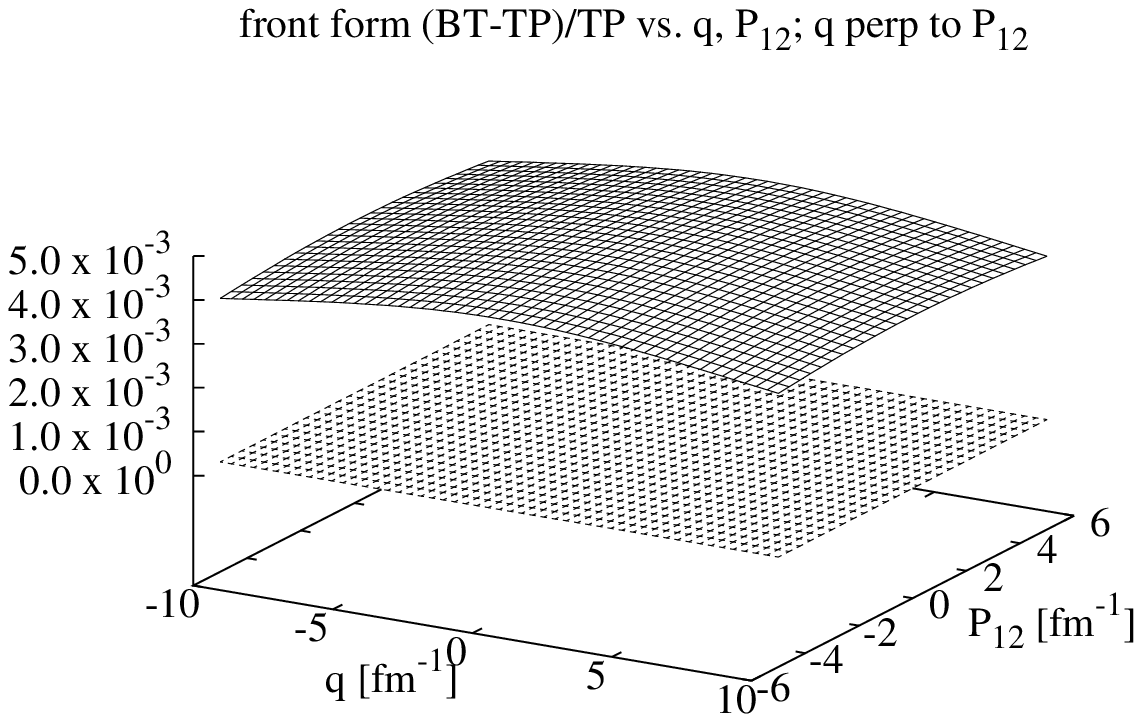}
\caption[Short caption for figure 3]{\label{labelFig1}                          
Front form  - $p_{12}\perp q$                                                   
}
\end{center}
\label{fig.1}
\end{minipage}
\begin{minipage}[t]{6.2cm}
\begin{center}
\includegraphics[width=5.9cm,clip]{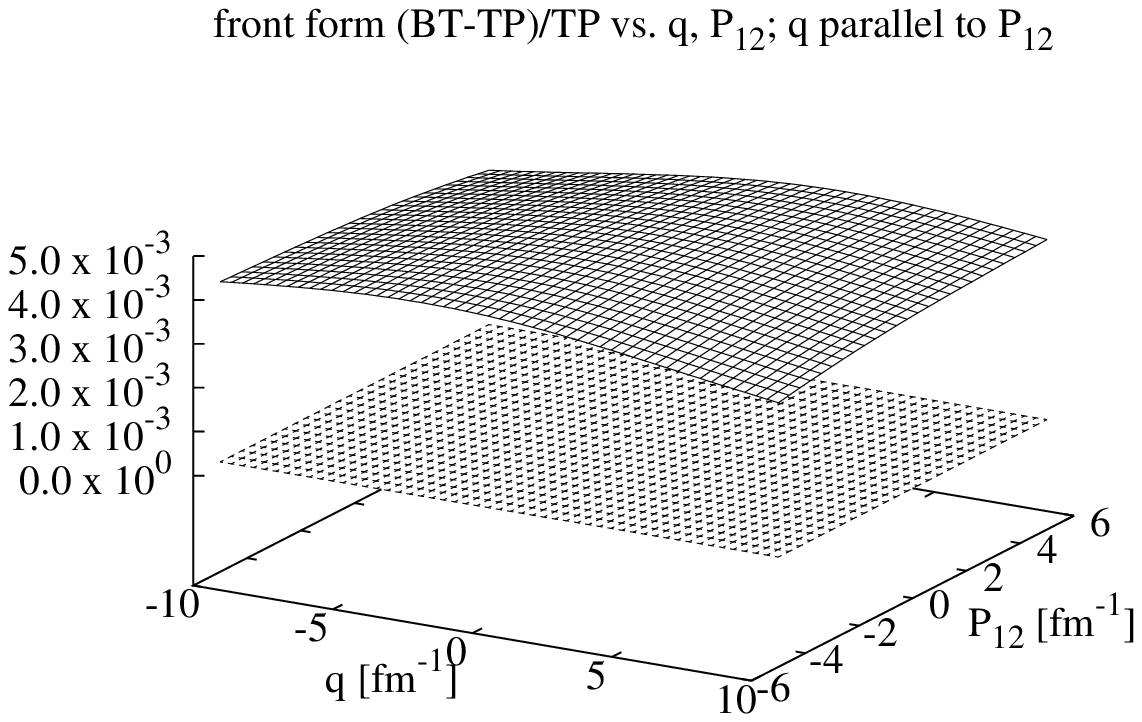}
\caption[Short caption for figure 4]{\label{labelFig2}                          
Front form  - $p_{12} \Vert  q$                                                 
}
\end{center}
\end{minipage}
\label{fig.2}
\end{figure}

\begin{figure}
\begin{minipage}[t]{6.2cm}
\begin{center}
\includegraphics[width=5.9cm,clip]{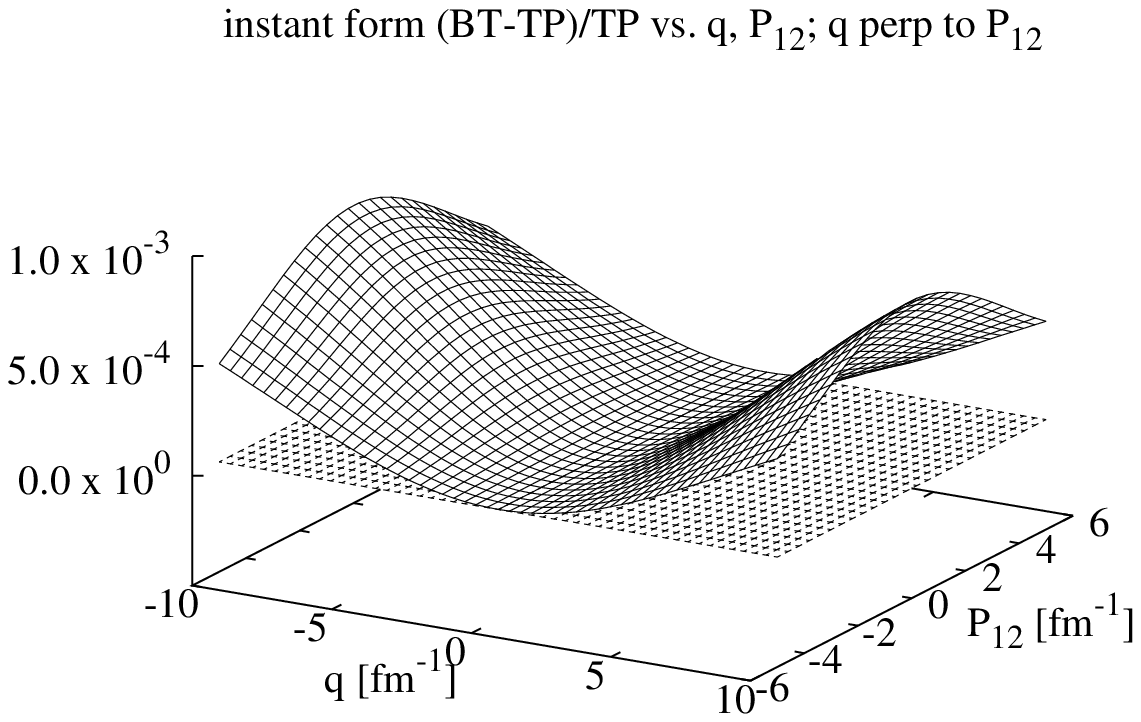}
\caption[Short caption for figure 3]{\label{labelFig3}                          
Instant form  - $p_{12}\perp q$                                                   
}
\end{center}
\label{fig.3}
\end{minipage}
\begin{minipage}[t]{6.2cm}
\begin{center}
\includegraphics[width=5.9cm,clip]{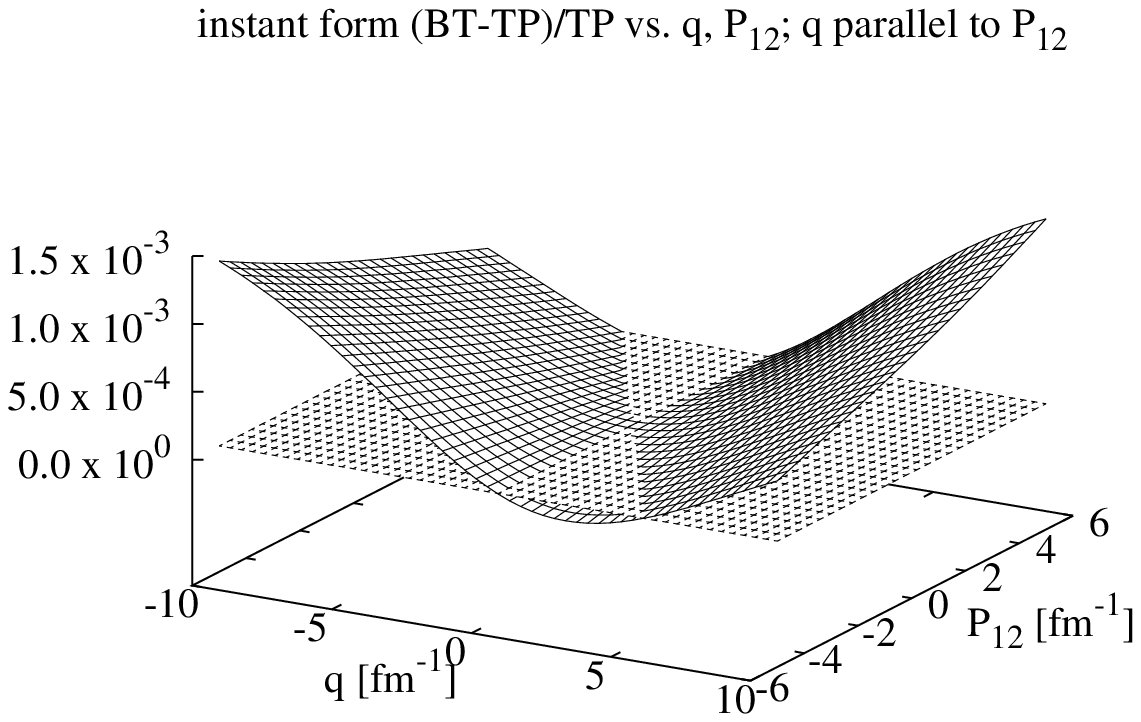}
\caption[Short caption for figure 4]{\label{labelFig4}                          
Instant form  - $p_{12} \Vert  q$}
\end{center}
\end{minipage}
\label{fig.4}
\end{figure}

\begin{figure}
\begin{minipage}[t]{6.2cm}
\begin{center}
\includegraphics[width=5.9cm,clip]{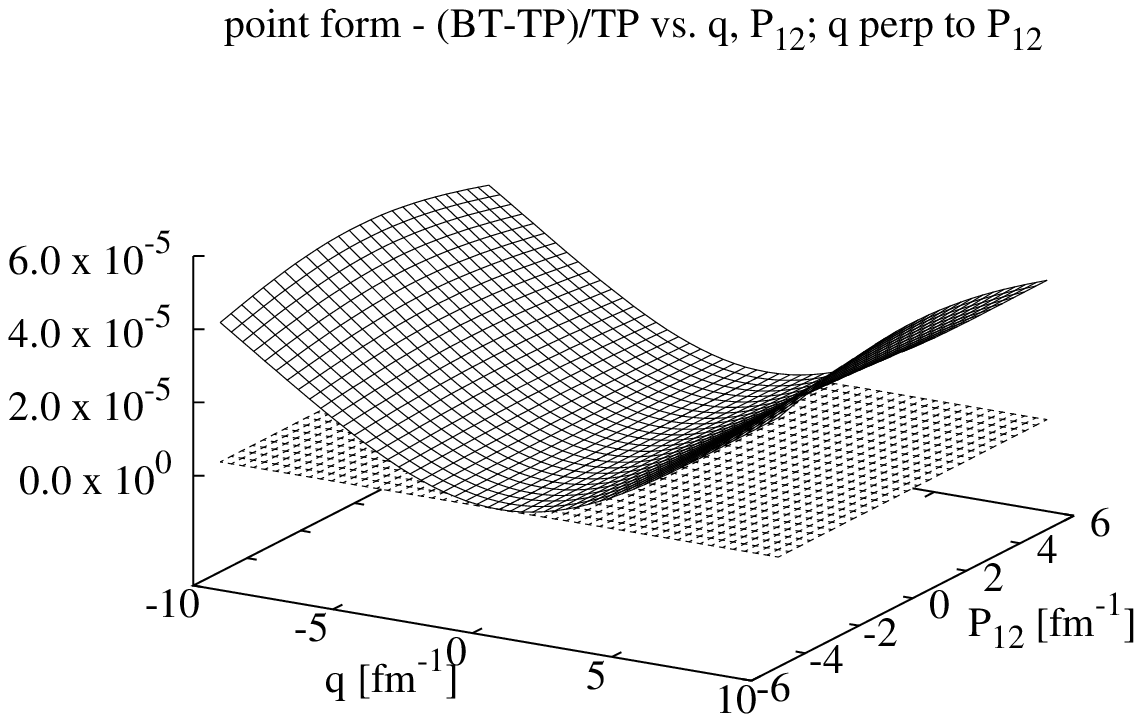}
\caption[Short caption for figure 6]{\label{labelFig5}                          
Point form  - $p_{12}\perp q$                                                   
}
\end{center}
\label{fig.5}
\end{minipage}
\begin{minipage}[t]{6.2cm}
\begin{center}
\includegraphics[width=5.9cm,clip]{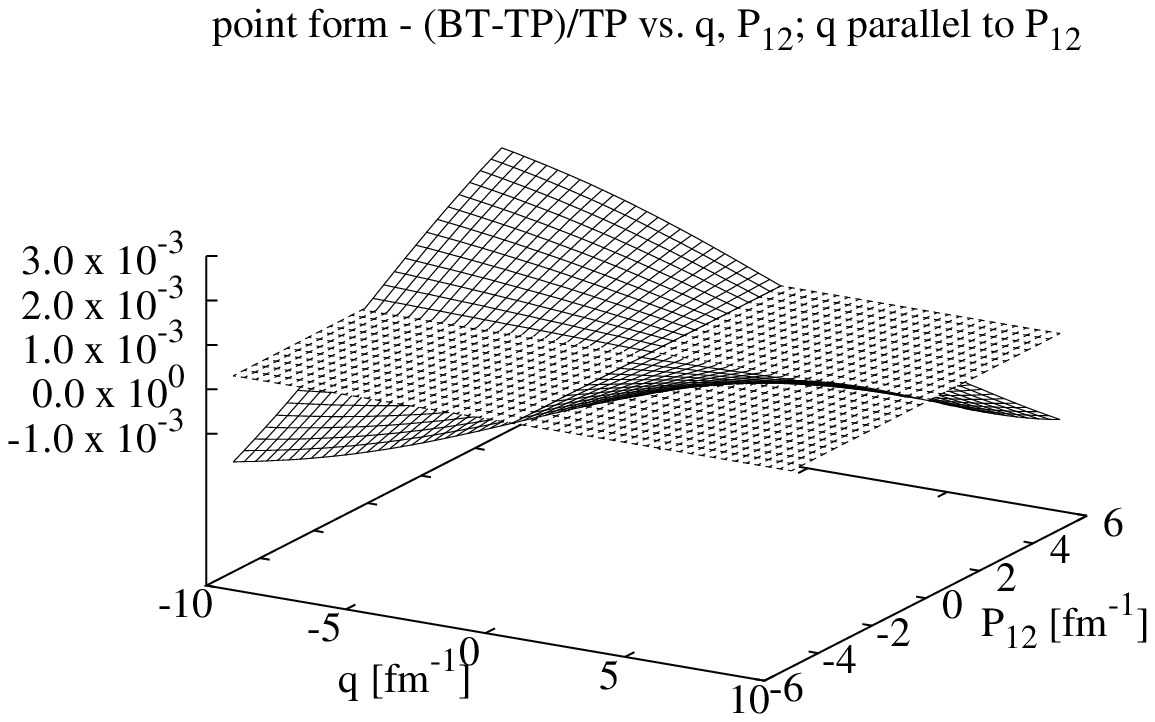}
\caption[Short caption for figure 6]{\label{labelFig6}                          
Point form  - $p_{12} \Vert  q$                                                 
}
\end{center}
\end{minipage}
\label{fig.6}
\end{figure}

\end{document}